\documentclass[usenatbib, usegraphicx]{mn2e}

\title[The galaxy population of the $z=1$ cluster of galaxies MG2016+112]{The galaxy population of the $z=1$ cluster of galaxies MG2016+112\thanks{Based on observations made with ESO Telescopes at the Paranal Observatory under programmes ID 63.O-0379 and 65.O-0666}}

\author[S. Toft, G. Soucail \& J. Hjorth]
       {S. Toft,$^1$\thanks{E-mail: toft@astro.ku.dk},
	G. Soucail$^2$ and J. Hjorth$^1$ \\
 $^1$Astronomical Observatory, University of Copenhagen, Juliane Maries Vej 30, 
DK-2100 Copenhagen \O, Denmark \\ $^2$ Observatoire Midi-Pyr\'{e}n\'{e}es, Laboratoire d'Astrophysique, UMR 5572, 14 avenue E. Belin, F--31400 Toulouse, France }

\pagerange{\pageref{firstpage}--\pageref{lastpage}}
\pubyear{2002}

\usepackage{natbib}
\newcommand{\beas}{\begin{eqnarray*}}
\newcommand{\eeas}{\end{eqnarray*}}
\newcommand{\mnras}{MNRAS}
\newcommand{\apj}{ApJ}
\newcommand{\apjs}{ApJS}
\newcommand{\apjl}{ApJL}

\newcommand{\aap}{A\&A}
\newcommand{\aaps}{A\&AS}

\newcommand{\aj}{AJ}

\newcommand{\pasj}{PASJ}
\newcommand{\nat}{Nature}

\begin{document}

\maketitle
\label{firstpage}

\begin{abstract}
A photometric redshift analysis of galaxies in the field of the wide-separation gravitational lens MG2016+112 reveals a population of  69
galaxies with photometric redshifts consistent with being in a
cluster at the redshift of the giant elliptical lensing galaxy $z=1.00$.
The $Ks$-band luminosity function of the
cluster galaxies is well represented by the Schechter function with a 
characteristic magnitude $K^*_s=18.90_{-0.57}^{+0.45}$ and faint-end slope  $\alpha=-0.60^{+0.39}_{-0.33}$, consistent with what is expected for a passively evolving population of galaxies formed at high redshift, $z_f>2$.
From the total $Ks$-band flux of the cluster galaxies and a dynamical estimate of the total mass of the cluster, the restframe $Ks$-band mass-to-light ratio of the cluster is derived to be 
$M/L_{Ks}=27^{+64}_{-17}\, h_{50}\,(M/L_{Ks})_{\sun}$, 
in agreement with the upper limit  derived earlier from 
Chandra X-ray observations and the value derived locally in the Coma cluster.
The cluster galaxies span a red sequence with a considerable scatter
in the colour-magnitude diagrams, suggesting that they contain young stellar populations in addition to the old populations of main-sequence stars that dominate the $Ks$-band luminosity function. This is in agreement with spectroscopic observations which show that 5 out of the 6 galaxies in the field confirmed to be at the redshift of the lensing galaxy have emission lines.
The projected spatial distribution of the cluster galaxies is filamentary-like rather than centrally concentrated around the lensing galaxy, and show no apparent luminosity segregation. A handful of the cluster galaxies show evidence of merging/interaction.  
The results presented in this paper suggest that a young cluster of galaxies is assembling around MG2016+112.

\end{abstract}


\section{introduction}

\subsection{Galaxy evolution in clusters}
The predominance of early-type galaxies in clusters compared with
the field is a clear sign that the evolution of galaxies depends on
environment, but the physical mechanism at work, in particular the
question of when and how the most massive galaxies, the giant
ellipticals, formed is a topic of ongoing theoretical and
observational research. 

The traditional ``monolithic'' elliptical galaxy formation scenario proposed by \cite{eggen}
 postulates a single burst of star formation at high redshift, followed by passive stellar evolution.

Early type galaxies in nearby and intermediate redshift clusters are observed to form a tight colour-magnitude relation
\citep{bower92,aragon-salamanca93,gladders98,SED98}, or  ``red sequence''
which when combined with the close correlation between galaxy mass and
metallicity implied by the Mg$_2$-$\sigma$ relation \citep{bender93}
is explained naturally by a single, early episode of star formation
($z_f>2$). 
If more massive galaxies are more efficient at retaining supernova
ejecta, they will have higher metallicities, and therefore redder
colours \citep{arimoto87,kodama97}.
 
There is however many pieces of evidence suggesting that the evolution
of the entire cluster galaxy population is not as simple as expected
from the monolithic galaxy formation scenario. 
The fraction of blue galaxies in clusters increase rapidly with
redshift \citep{butcher78,butcher84,rakos95}. High resolution imaging
and  spectroscopic studies show that a significant fraction of
intermediate redshift blue galaxies are late-type spirals
with emission lines and/or post-starburst signatures
(\cite{dressler92,dressler99,poggianti99} and that the cluster S0
population is far less abundant at $z\sim0.5$ than today. It has been proposed that
these blue star-forming galaxies are the progenitors of cluster S0 galaxies 
 \citep{dressler97,vandokkum98,vandokkum00}. In agreement with this
explanation of the Butcher-Oemler effect, some S0s in the present
epoch appear to contain younger stellar populations \citep{kuntschner98}.

Hierarchical structure formation models present a very different view
of galaxy evolution in which galaxies assemble by merging of smaller
stellar systems over a wide redshift range. In this scenario,
the evolution of elliptical galaxies are governed
by the time at which the bulk of the stellar populations are formed,
the era when the majority of mergers took place, and the amount of new
star formation induced during each merger event.
If the formation of the stars in an elliptical
galaxy takes place over a broad cosmic time interval, then
ellipticals at any redshift should exhibit a wide range of
mass-weighted stellar ages. If however most stars in present day
cluster ellipticals were formed in smaller disks at high redshift
$z\gg1$ and if little additional star formation took place in
subsequent mergers, then the end product ellipticals could appear to
be old and approximately coeval even if the bulk of merging took place
relatively late. The hierarchical models of \cite{kauffmann98}
reproduce the colour-magnitude relation of ellipticals because more
massive galaxies are formed from merging of systematically more massive
progenitors, which retain more metals when forming stars.

Discriminating between the two formation scenarios described above is
difficult, even when studying galaxies in clusters at high
redshift. An increasing number of observations 
\citep{stanford97,ellis97,SED98,rosati99,nakata01}  show that the
evolution of the ``red sequence'' to redshifts $z\sim1$ is consistent
with a high formation redshift and a subsequent passive evolution
favored by monolithic collapse, but none of them
can exclude the possibility that cluster ellipticals formed from
mergers of smaller galaxies as long as the bulk of star formation 
took place at much larger redshifts and there was little star
formation in the subsequent merging process.

One way to distinguish between the two scenarios is by studying the
evolution of the mass function of the cluster galaxies.
Hierarchical galaxy formation models predicts the mass assembly of
galaxies to take place over a long time scale ($z\ll2$) and to be
quite decoupled from star formation \citep{kauffmann98b}.
If this picture is true, a strong evolution of the stellar mass
function  with redshift is expected. This evolution is best studied in
the $Ks$-band because it is relatively unaffected by ongoing/recent
star-formation and hence directly measures the growth of galaxy mass
\citep{kauffmann98b}. 

Locally, the cluster galaxy luminosity function have been studied in
great detail both at optical and Near Infrared (NIR) wavelengths.  \cite{goto02}
derive the composite luminosity function of 204 ``Sloan Digital Sky
Survey Cut \& Enhance Galaxy Cluster Catalog'' in the redshift range
range $z=0.02$ to $z=0.25$ in the five SDSS bands $u^{*}$, $g^{*}$,
$r^{*}$, $i^{*}$
and $z^*$, and find  that the  
faint-end slope of the luminosity function becomes flatter toward the
redder wavebands, consistent with the hypothesis that the cluster
luminosity function has two distinct underlying populations; a
population of bright ellipticals with a gaussian-like luminosity
distribution that dominate the bright end, and a population of faint
blue star-forming galaxies with a steep power-law like luminosity
distribution, that dominate the faint-end.
\cite{depropris98coma} derive the NIR (H-band) luminosity function
of the Coma cluster galaxies, and find results consistent with the above
picture of  a
population of bright red galaxies and large population of faint blue
dwarf galaxies. 

HST imaging of high redshift cluster galaxies have revealed a population of luminous ``red'' mergers in high redshift clusters
\citep{vandokkum00,vandokkum01}. 
The existence of these galaxies which follow the same colour-magnitude relation as the
cluster ellipticals (but with a slightly larger scatter) 
supports the hierarchical galaxy formation scenario.

The bright (massive) end of the luminosity function is expected to
steepen and shift to fainter magnitudes at high redshift, as the old
bright cluster galaxies begin to break up into building blocks. 

In contrast to the bright cluster galaxies, the faint cluster galaxies
seem to have a much greater diversity of star formation histories
\citep{kodama01}. The Butcher-Oemler effect in intermediate redshift
is an example of this. 
If this progression of activity to lower-mass galaxies as the universe
ages is a consequence of the faint galaxies forming at later cosmic
times than their massive counterparts, then distant clusters should
have a luminosity function with a declining faint-end slope in
contrast to rising faint-end slope of local clusters \citep{depropris98coma}.
  
The evolution of the $K$-band luminosity function in the redshift range
$z=0.1$-$0.9$ has been studied by \cite{depropris99}. The
evolution of the characteristic magnitude $K^*$ of the galaxies,
is found to be consistent with what is expected for a passively evolving population of
galaxies formed at $z_f>2$. The data
is not deep enough to constrain the faint-end slope 
which is fixed at $\alpha=-0.9$ (the value derived in the H-band for
the Coma cluster). The authors find no significant dependence on the
X-ray luminosity of the clusters.    

\cite{nakata01} extend the study to $z=1.2$ by deriving the $K$ band
luminosity function for the cluster around the radio galaxy 3C324. 
These authors also choose to fix the faint-end slope at  $\alpha=-0.9$
and find consistency with what is expected for passively evolving
stellar populations formed at $z_f>2$.

The suggested passive evolution of $K^*$ with redshift to $z=1.2$ is a nice
confirmation of the passively evolving stellar population found from
other methods (such as the colour-magnitude relation), but does not
help discriminate between the monolithic and hierarchical formation
scenarios, since no steepening of the bright end of the luminosity
function (a deficit of the brightest galaxies) has been observed and
no attempt has been made to measure the faint-end slope.

\subsection{Previous observations of the MG2016+112 system }
MG2016+112 is a wide separation gravitational lens
first identified by \cite{lawrence84} to be comprised of
at least three lensed images of an active galactic nucleus (AGN) at 
redshift $z=3.269$. The primary lensing object is a giant elliptical
galaxy at $z=1.004$ (hereafter ``the lens galaxy'', \cite{schneider86,soucail01,koopmans02}).
Wide separation gravitational lens systems have in many cases been found to
contain galaxy clusters or groups that contribute significantly to
the large separation angles.  
A nice property of clusters discovered in this way is that they are mass selected, and therefore not biased toward the old relaxed systems that clusters selected in X-rays or in the optical/NIR are (since their selection depends on the presence of thermalized gas and/or central concentrations of luminous old red galaxies).

\cite{hattori97} reported the detection of resolved X-ray emission
at $z\sim 1$ centered on the lens galaxy. This result was
based on the detection of an X-ray line at 3.35 keV in the
ASCA  spectra of the system, which was interpreted as Fe K emission
originating from a cluster at $z\sim1$, and on the detection
of  ``diffuse'' X-ray emission 
near the giant elliptical in ROSAT HRI observations.  
Infrared and optical observations of the field revealed no excess of
galaxies around the giant elliptical lens
\citep{schneider85,langston91,lawrence93}. This led to the 
suggestion of a ``dark cluster'' associated with the lens, in which a
cluster sized dark matter and hot gas overdensity exists with few optically
bright galaxies ($M/L_V > 2000M_{\sun}/L_{\sun}$). This conclusion was
puzzling as the observed iron line in the X-ray spectra indicated a
near solar metallicity for the cluster, which implies a long history
of star-formation in the region.
\cite{benitez99} re-analyzed the ROSAT HRI data and found a faint
elliptical X-ray source which was coincident with a red galaxy
overdensity. Deep $V$,$I$ and $K$ band Keck imaging revealed a red sequence
of galaxies, with colours consistent with being at redshift $z\sim1$.

\cite{soucail01} performed a spectroscopic survey of 44 galaxies in
the field and found an excess of 6 red galaxies securely identified to
be at $z\sim 1$, though not very centrally concentrated around the
giant elliptical galaxy. For these six galaxies a mean redshift of
$z=1.005$, a velocity dispersion of
$\sigma=771^{+430}_{-160}$km\,s$^{-1}$, a dynamical mass estimate of $2.8\times10^{14}h_{50}M_{\sun}$ and a V-band mass-to-light ratio
of $M/L_V=215^{+308}_{-77}h_{50}M_{\sun}/L_{\sun}$ was derived.

It is interesting to note that a high fraction of the $z=1$ galaxies (5 out of 6) have emission lines in their spectra, indicating the presence of young stellar populations.  This is a much larger fraction than in low redshift clusters which are dominated by old quiescent galaxies with no emission lines, and also much higher than in other high redshift ($z \ge 1$) clusters, where the fraction of emission line galaxies is found to be 10-15\%  \citep{stanford97,rosati99}.

\cite{clowe01} performed a weak lensing analysis of the field and
detected a mass concentration 64{\arcsec} northwest of the lensing
galaxy, and were able to rule out (at the $2\sigma$ level) a mass peak
centered on the lensing galaxy of the size suggested by the
\cite{hattori97} X-ray observations.

\cite{chartas01} recently acquired a 7.7 ksec observation of the system
with Chandra. These observations can account for all the X-ray emission
as originating from the lensed images and additional  X-ray point
sources in the field, and place an $3\sigma$ upper limit on the 2-10
keV flux and luminosity of diffuse cluster emission of 1.6 $\times$ 10$^{-14}$ ergs
s$^{-1}$ cm$^{-2}$ and 1.7 $\times$ 10$^{44}$ ergs s$^{-1}$,
respectively, and an estimated upper limit on the mass
$M_{500}<4.5\times10^{14}h_{50}M_{\sun} $ and mass-to-light ratio
of $M/L_V <190M_{\sun}/L_{\sun}$.

Rather than being an X-ray luminous ``dark-cluster'', MG2016+112 thus
turned out to be a spectroscopically confirmed cluster of galaxies,
detected neither in X-rays nor in weak lensing maps.

In this paper we revisit the system through deep NIR and optical
observations. We identify 69 galaxies in the field with photometric
redshifts consistent with being cluster galaxies at redshift $z=1$ and
derive their colour-magnitude sequences and their luminosity
function.

The outline of the paper is as follows:
In section \ref{observations} we describe the observations and data
reduction. In section \ref{photometry} we describe the construction of a
multi colour photometric catalog of galaxies in the field (object
detection, matching, photometry, star/galaxy separation). In section
\ref{photoz} we discuss the photometric redshift analysis. The results are
presented in section \ref{results} and summarized in Section
\ref{summary}.

Throughout this paper we assume a flat $\Omega_{\Lambda}=0.3$,
$\Omega_{\Lambda}=0.7$, $h_0=0.70$ cosmology unless otherwise noted.

\section{Observations and data reduction}
\label{observations}
The observational basis of this work consists of 
very deep NIR  $J$ and $Ks$
band observations carried out with the ISAAC instrument (proposal
63.O-0379) in service mode between May and July 1999, and moderately
deep optical $B$, $V$, $R$, and  $I$ band observations carried out
with the FORS1 instrument (proposal 65.O-0666) in service mode in July
2000. Both FORS1 and ISAAC are mounted on the VLT/ANTU
telescope. 
ISAAC employs a
$1024^2$ Hawaii array, with a pixel scale of $0\farcs147$ and a field
of view (FOV) of $2\farcm5\times2\farcm5$. 
FORS1 employs a $2048^2$ CCD (TK2048EB4) which with the high resolution
collimator used for the present observations has a pixel scale of
$0\farcs1$ and a FOV of $3\farcm4\times3\farcm4$.

The data was originally obtained for other purposes than studying the
cluster galaxy population. The NIR dataset was designed to perform a
weak lensing analysis on the background galaxies. The optical dataset was designed to study the central bright gravitationally lensed QSO and is therefore not deep enough to detect the fainter cluster galaxies. The derived magnitudes (or upper limits) however serve as powerful input for the photometric redshift analysis which separates the cluster galaxies from the foreground and background field galaxy population.
All observations were performed in excellent seeing conditions. 
The observing log is given in Tab.~\ref{data}.

\begin{table}
\caption{Log of the VLT/UT1 observations}
\label{data}

\begin{tabular}{@{}lcccccc}\hline\hline
Instrument      &Filter &NExp   &Exp&Seeing &Completeness\\ \hline
ISAAC	&$Ks$	&275 	&60 &0\farcs4 & 22\fm0 \\ 
ISAAC	&$J$	&70	&120&0\farcs4 & 23\fm0\\ \hline
FORS1	&$I$	&2     	&200&0\farcs4 & 24\fm0\\ 
FORS1	&$R$	&2	&150& 0\farcs4& 24\fm5\\	
FORS1	&$V$	&4	&180& 0\farcs4& 26\fm0\\	
FORS1	&$B$	&5	&300& 0\farcs5& 27\fm0 \\ \hline
\end{tabular}
{\small{NExp: number of exposures, Exp: exposure time in seconds}}
\end{table}

\subsection{Reduction and calibration of NIR data}
The deep near-infrared imaging consists of 70 images of 2 minutes
integration time (2 hours 20 minutes)  obtained in $J$ in 3
independent sequences, and 275 images of 1 minute each (4 hours 35
minutes) obtained in $Ks$ in 5 sequences. Some of them were taken
in photometric conditions while others not, so subsequent re-scaling
to the photometric ones was applied in the final combination. 
Data reduction was performed with the ``eclipse'' software
package \citep{eclipse} and IRAF routines. Flat-fields were produced
from series of twilight sky flats. 
To take into account the rapidly varying NIR background the sky value in each pixel was calculated as a running median of the pixel value in the exposures taken immediately before and after a given exposure.

To compensate for the dithering pattern and optical distortion the individual frames were transformed to a reference frame using a 3rd order geometrical transformation calculated from the position of 30-40 stars in each frame. Finally, the images were averaged with some clipping to remove residuals defects. 
Photometric calibration was obtained through the observations of
standard calibration stars during the photometric nights of the
programme.

\subsection{Reduction and calibration of optical data}

The individual optical data frames were bias subtracted and
flat fielded using the standard ESO pipeline and cleaned for cosmic rays
using Laplacian edge detection \citep{laplace}. In some
of the frames the background was found to have large scale gradients
over the field. To
properly account for this we ran all the frames through the SExtractor
software \citep{bertin} with options set to save a full resolution
interpolated background map, which was then subtracted from the
science frames (large scale background variation are well modeled by SExtractor). 
Cosmetic defects in areas with no objects were patched by interpolation of the
background in surrounding annuli. This was necessary since some
wavebands had too few exposures for deviant pixel rejection. 
Finally the individual frames were geometrically transformed to the
same reference frame and combined using the IRAF tasks {\emph{geoform}},
{\emph{geotran}} and {\emph{imcombine}}.

The combined frames were calibrated using the atmospheric extinction
and color corrected zero-points quoted on the ESO webpage, and
corrected for Galactic extinction using the parameterization of the
Galactic extinction law of \cite*{CCM}, assuming the mean Galactic
value for the total-to-selective extinction in the V band $R_V=\left<R_V\right>_{MW}=3.1$ and
E(B$-$V)=0.22 (from interpolation of DIRBE/IRAS dust maps,
\citealt{schlegel}). Finally, all the frames were WCS calibrated using
the USNO-A2.0 catalog \citep{USNO}.

\section{Object detection, Photometry and Matching}
\label{photometry} 
Object detection was done in the $Ks$ band, using the SExtractor
software \citep{bertin} with default parameter settings.
We searched for the detected objects in the remaining frames using the
SExtractor ASSOC option, which assumes that the frames being
matched have been registered to the same pixel coordinate system. 

The transformations of the combined images to the $Ks$-band image were
derived using the IRAF tasks {\emph{geoform}}. 

Transforming the optical images to the $Ks$-band image includes a
rebinning (due to the different pixel scales of the two instruments)
in addition to the distortion transformation. To avoid this image
degradation, as well as other potential problems associated with 
non-linear image transformations, such as blurring of
persistent bad pixels and cosmic ray events, corruption of the background
statistics etc. we transformed the $Ks$-band object list xy-pixel
coordinates to match the reference frames of the other images
using the IRAF task {\emph{geoxytran}}, rather than transforming the
images themselves.   
In this way the photometry is performed  on the original,
untransformed images.  

For the total $Ks$ band magnitudes of the objects we used  SExtractors
{\emph{mag\_best}} which has been shown to be accurate to about
$0\fm05$ \citep{smail2001}  and colours were derived in fixed apertures
of approximately 12 kpc at $z=1$ (10 pixels in the ISAAC images, 15 pixels in
the FORS1 images).
  
\subsection{Completeness}
To estimate the completeness magnitude of the observations we ran
SExtractor object detection independently on all the frames. 
In Fig.~\ref{complete} we plot the number of detected objects as a
function of $Ks$ and $J$ band magnitude.  From visual inspection of the turnover magnitudes of the distributions in  Fig.~\ref{complete} we estimate that
the  $Ks$ band observations are complete to about 22\fm0
and the $J$ band observations are complete to about 23\fm0. The
estimated completeness magnitudes of all the wavebands, including the
optical are listed in Tab.~\ref{data}.
 
\begin{figure}
\setcounter{figure}{0}
\resizebox{\hsize}{!}{{\includegraphics{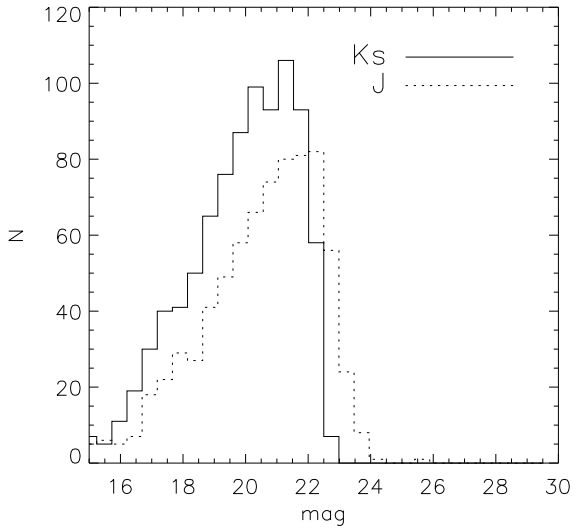}}}
\caption{Binned representation (binsize is $0\fm5$) of the number of
detected objects as a function of magnitude in the $Ks$ (full curve) and $J$ band (dashed curve). 
The data are estimated to be complete to $Ks$=22\fm0 and $J$=23\fm0 
}
\label{complete}
\end{figure}

\subsection{Star/galaxy separation}
SExtractor assigns to each object a parameter {\emph{class\_star}}
 which ranges from 0 to 1. When  the FWHM is well
determined stars obtain values close to 1, while extended objects
obtain values close to 0. A main concern is contamination of
the galaxy sample by stars, especially since some of them (e.g. M stars) would have the same colours as early-type galaxies at the
cluster redshift. We therefore adopt a very conservative threshold in
the {\emph{class\_star}} parameter of 0.05. Objects with {\emph{class\_star}} larger than
this value in the $Ks$ band or $J$ band are labeled stars.
Now we can construct a multi colour catalogue of galaxies consisting
of objects which (i) have $Ks\le22$ (ii) are detected in at least two
bands and (iii) are classified as a galaxy by means of the {\emph{class\_star}} parameter.

\section{Photometric Redshift Analysis}
\label{photoz}
To separate the cluster galaxies from the foreground and background
galaxies 
we estimated their redshifts using the  public {\emph{hyperz}}
photometric redshift code of \cite{hyperz}. The code uses the GISSEL98 (Galaxy Isochrone Synthesis Spectral Evolution Library) spectral
evolution library of \cite{bruzual} to build synthetic spectra
with 7 different star formation histories which match the observed
properties of local galaxies from E to Irr type: a constant star-forming rate and six exponentially decaying star formation rates
with time scales from 1-30 Gyr. The models assume solar metallicity
and a Millar-Scalo IMF, and internal reddening is considered using the
\cite{calzetti} model with $A_V$ varying between 0\fm0 and 1\fm2.
Objects not detected in a waveband are assigned zero flux (with an error
corresponding to the limiting magnitude) rather than just ignored in the fit. This ``upper limit''
is important for the photometric redshift determination and is the
reason for including the rather ``shallow'' optical data in the
analysis.
Fig.~\ref{photoz1} shows a  histogram of the derived photometric
redshifts of all the galaxies in the field.
There is a distinct peak at redshift $z_{phot}\sim1$, and quite a few
galaxies with $z_{phot}\la0.2$.
\begin{figure}
\setcounter{figure}{1}
\resizebox{\hsize}{!}{{\includegraphics{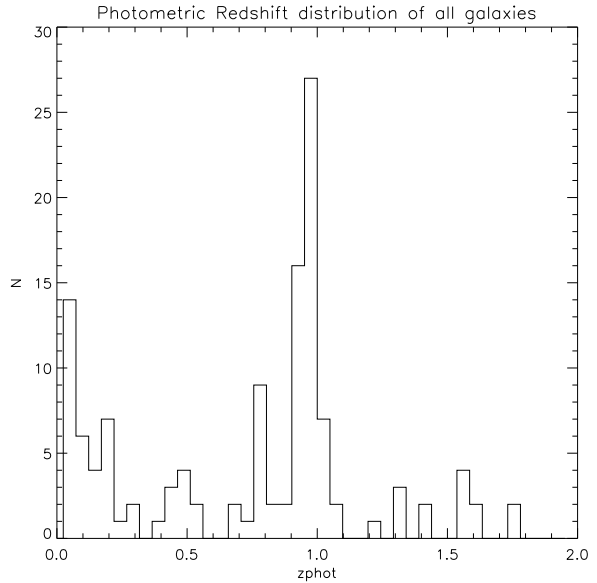}}}
\caption{Photometric redshifts distribution of the galaxies in the
field of MG2016+112 derived using {\emph{hyperz}}. }
\label{photoz1}
\end{figure} 

\begin{figure}
\setcounter{figure}{2}
\resizebox{\hsize}{!}{{\includegraphics{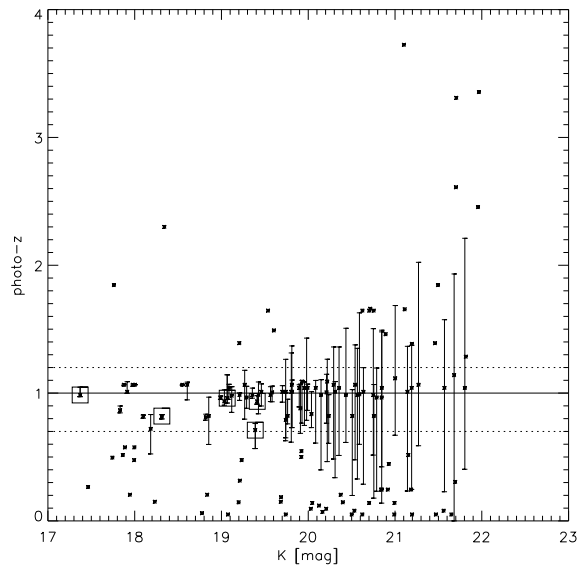}}}
\caption{Photometric redshifts of all galaxies in the field of
MG2016+122 versus their $Ks$ band magnitudes. The error bars are 1$\sigma$ from {\emph{hyper-z}}. The squares mark galaxies
which have been spectroscopically confirmed to be at redshift $z=1$ \citep{soucail01}}
\label{photoz2}
\end{figure} 

\begin{figure*}
\setcounter{figure}{3}
\resizebox{\hsize}{!}{{\includegraphics{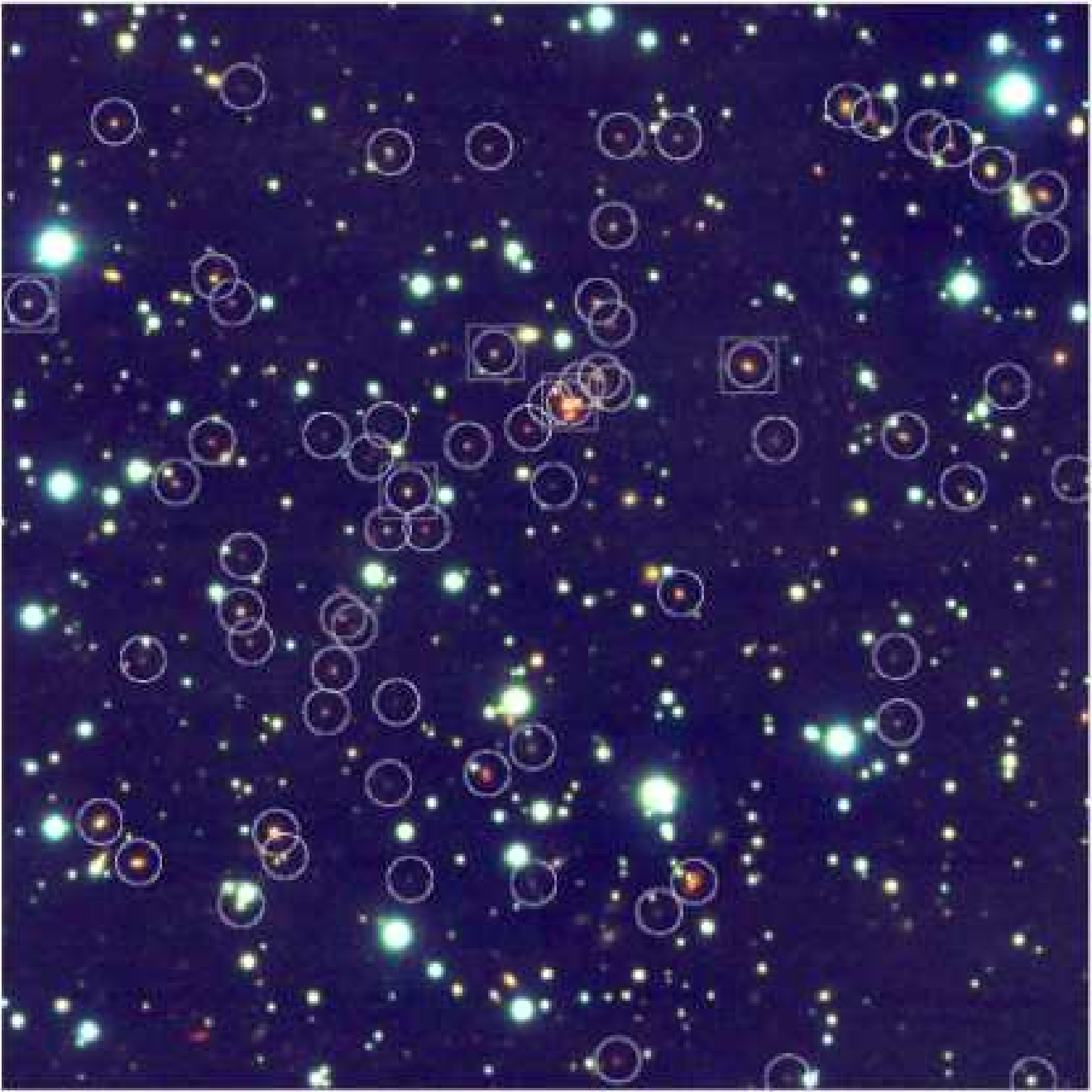}}}
\caption{Composite $Ks$, $Js$, $I$-band image of the field of MG2016+112. The field is 2.5\arcmin $\times$2.5\arcmin. The pixel scale is $0\farcs147\,$pix$^{-1}$. The circles mark the
69 ``photometric members'' of the cluster at $z=1$. The squares mark the ``spectroscopic members'' of the cluster. North is up, East is to the left. }
\label{distrib1}
\end{figure*}

\begin{figure}
\setcounter{figure}{4}
\resizebox{\hsize}{!}{{\includegraphics{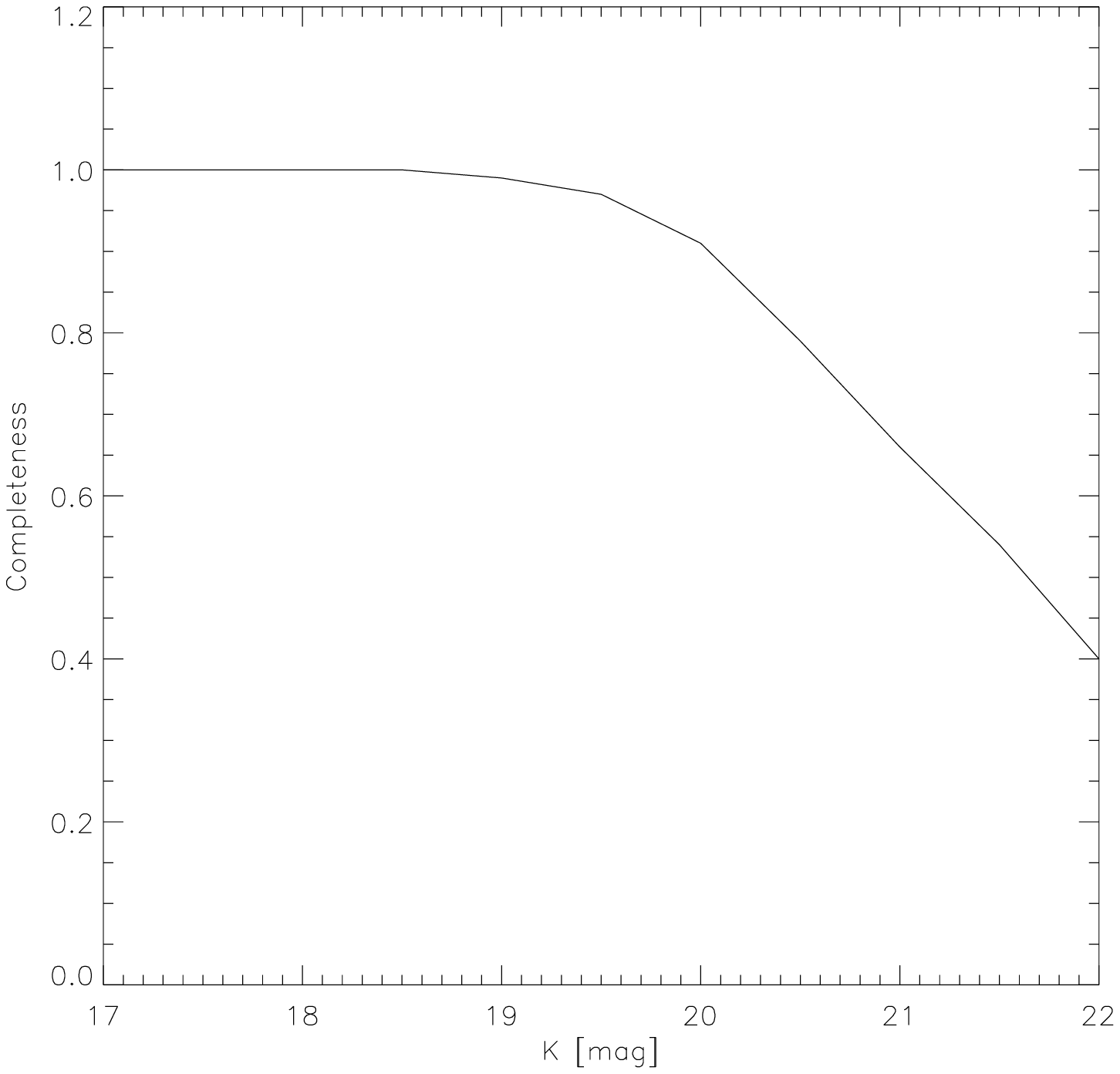}}}
\caption{Simulated fraction of redshift $z=1$ galaxies with derived photometric
redshifts in the range $0.7<z_{phot}<1.2$ as a function of $Ks$ band magnitude  }
\label{complfunct}
\end{figure} 

In Fig.~\ref{photoz2} we plot the  derived photometric redshifts of
all the galaxies in the field versus their $Ks$-band magnitude. The open
squares mark the galaxies which have been spectroscopically confirmed
to be at redshift $z=1$ (hereafter spectroscopic members, \cite{soucail01}). 
The full line marks the
redshift of the lensing galaxy at $z=1$ while the dashed lines are the limits
$z_{phot}=0.7$ and $z_{phot}=1.2$ that encompasses all the galaxies in the peak
around $z_{phot}=1$, including the 5 spectroscopically confirmed
ones.     
Fig.~\ref{photoz2} reveals that  most of the $z_{phot}\la0.2$ are faint
$Ks>20$  galaxies which may have been misclassified as dusty low redshift
galaxies due to poor photometry.  

We classify the 69 galaxies within the limits $0.7\le z_{phot}\le1.2$ as probable cluster galaxies (hereafter photometric members). 
The choice of limits is a trade off  between maximum cluster galaxy completeness and minimum contamination from field galaxies.
We choose the lower limit to be $z_{phot}>0.7$ rather than e.g. $z_{phot}>0.8$ to avoid excluding spectroscopic members, and other cluster galaxies with similar colours. Young stellar populations in the cluster galaxies can influence their broadband colours and derived photometric redshifts due to the discrete nature of the spectral library used by {\emph{hyperz}}. Since a large fraction of the spectroscopically confirmed cluster galaxies show evidence of recent star formation it is likely that this is also the case for other cluster galaxies.
Some galaxies in the $0.7<z_{phot}<0.8$ interval may be truly foreground, but many could also be cluster galaxies at $z=1$ with recent star formation.
To test the stability of the photometric redshifts derived with {\emph{hyperz}} versus the photometric errors we did the following:
For each $\Delta m$=0.5 mag bins in the observed range of
magnitudes $Ks=16-22$, we generated a catalogue of 1000 galaxies at $z=1$
for the same bandpasses and limiting magnitudes as for the data of
MG2016+112.
The catalogue was generated using the {\emph{make\_catalog}} code in
the {\emph{hyperz}} package. The galaxies are randomly drawn from the 7
template spectral types (E to Im) used for estimating the photometric
redshifts.
We then apply {\emph{hyperz}} to the catalog to estimate the photometric
redshifts.
In Fig.~\ref{complfunct} we plot the fraction of the input ($z=1$)
galaxies that have derived photometric redshifts in the range
$0.7<z_{phot}<1.2$ as a function of $Ks$-band magnitude.
From this figure it can be seen that {\emph{hyperz}} recovers 100\% of the
galaxies down to magnitudes of $Ks=18.5$. At $Ks=20$ it recovers
about 90\% and at $Ks=22$ the recovery rate is down to 40\%.

The redshift evolution of the field galaxy luminosity function, and its cosmic variance is not known with sufficient accuracy to correctly incorporate the effects of pollution from field galaxies in the completeness function analysis.
Instead we will return to the issue of field contamination in Sec.~\ref{distrib} where we discuss the derived projected spatial distribution of the cluster galaxies, and in Sec.~\ref{LF} where we discuss their luminosity function.

\section{Results}
\label{results}
\subsection{Distribution of the cluster galaxies} 
\label{distrib}
The distribution of the cluster galaxies (photometric members) on the sky is not very centrally concentrated. This is illustrated in Fig.~\ref{distrib1}
where the overlay circles 
mark the photometric cluster members, and in Fig.~\ref{densitycontours} where the overlay represents contours of the smoothed density distribution of the photometric member galaxies. The contours are 3-30$\sigma$ above the background density which is taken to be the density of galaxies in the HDFN with $K_s<22$ and $0.7<z_{phot}<1.2$ \citep{fernandez-soto99}.         
The apparent morphology of the cluster is an elongated filamentary-like distribution similar to what is observed in other $z>1$ clusters (e.g. the Lynx clusters \citep{stanford97,rosati99} and the RDCS J0910+5422 cluster \citep{stanford02}) and consistent with what is expected for a ``dynamically young'' cluster  in the process of assembly.  
The high significance of the overdensity structure, compared to the HDFN indicates that most of the photometric members are $z=1$ cluster galaxies rather than random field galaxies that would be distributed more homogeneously across the field.  
Fig.~\ref{distrib2} illustrates that there is no apparent luminosity
segregation of the cluster galaxies  (i.e the brighter cluster
galaxies are not more centrally clustered than the fainter galaxies).
This is consistent with the picture of a young cluster where the smaller galaxies have not yet had time to merge and transform into the population of massive galaxies found in the center of relaxed clusters.

\begin{figure}
\setcounter{figure}{5}
\resizebox{\hsize}{!}{{\includegraphics{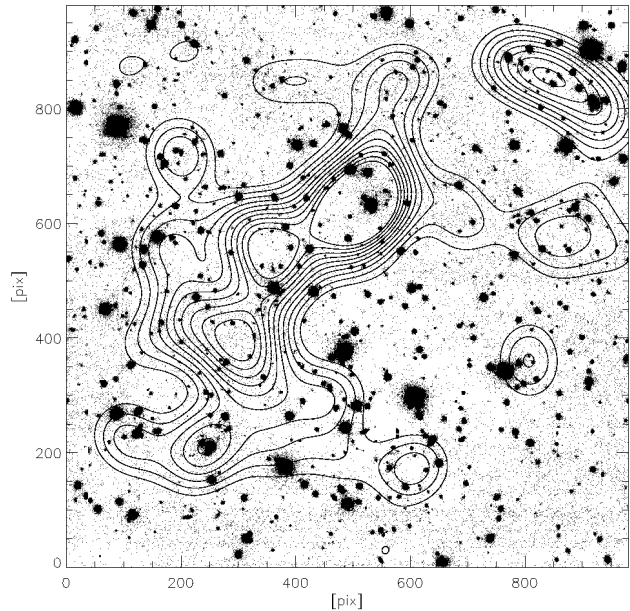}}}
\caption{Contours of the smoothed density distribution of photometric member galaxies overlayed the $Ks$-band image of MG2016+112. 
The contours are 3-30$\sigma$ above the background density which is taken to be the density of galaxies in the HDFN with $K_s<22$ and $0.7<z_{phot}<1.2$ \citep{fernandez-soto99}.
 The largest concentration of photometric member galaxies is centered on the central lensing galaxy, and is part of a ``loose'' elongated filamentary like structure. The pixel scale is $0\farcs147\,$pix$^{-1}$. North is up, East is to the left.  }
\label{densitycontours}
\end{figure} 

\begin{figure}
\setcounter{figure}{6}
\resizebox{\hsize}{!}{{\includegraphics{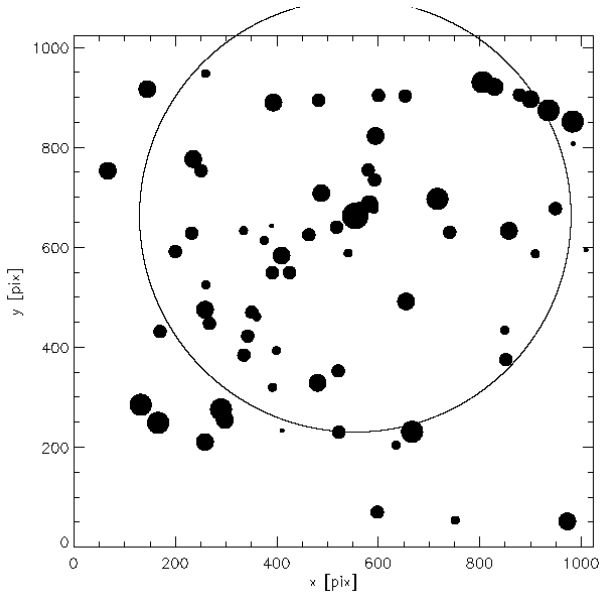}}}
\caption{Distribution of the ``photometric members'' of the cluster
at $z=1$. The area of the circles are proportional to the $Ks$-band
 brightness of the galaxies. The physical size of the radius of the
large circle is 0.5 Mpc. The pixel scale is $0\farcs147\,$pix$^{-1}$. North is up, East is to the left.   }
\label{distrib2}
\end{figure}

\subsection{Properties of the cluster galaxies}
A full morphological analysis of the cluster galaxies requires higher resolution imaging data, and is beyond the scope of this paper, but qualitative inspection of Fig.~\ref{distrib1} shows a wide variety of galaxies types:
elliptical, spirals and even three or four mergers.
Merging cluster galaxies are very rare in the cores of local relaxed clusters. The significant fraction of merging cluster galaxies found here, and in other high redshift clusters \citep{vandokkum00,vandokkum01} is direct evidence of the hierachical formation scenario.

\subsection{Colour-magnitude relation}
In Fig.~\ref{JK} and Fig.~\ref{RK} we plot the $J-Ks$ versus $Ks$ and $R-Ks$ vs $Ks$ colour-magnitude relations of galaxies in the field of MG2016+112.
Filled symbols mark galaxies within 0.5 Mpc of the lensing galaxy.  We label this region the ``inner region''. Open symbols mark  galaxies farther away than 0.5 Mpc from  the lensing galaxy. We label this region the ``outer
region''. The area of the outer region is approximately the same as the    
inner region.
Also indicated are predicted colour-magnitude relations at the cluster redshift ($z=1$) for passively evolving galaxies formed at $z_f$=2, 3 and 5 \citep{kodama97}.
This model assumes that the red sequence of early type galaxies in the colour-magnitude diagram is a pure metalicity sequence as a function of galaxy magnitude, and calibrate its zeropoint using the red sequence of early type galaxies in the Coma cluster. 

\begin{figure}
\setcounter{figure}{7}
\resizebox{\hsize}{!}{{\includegraphics{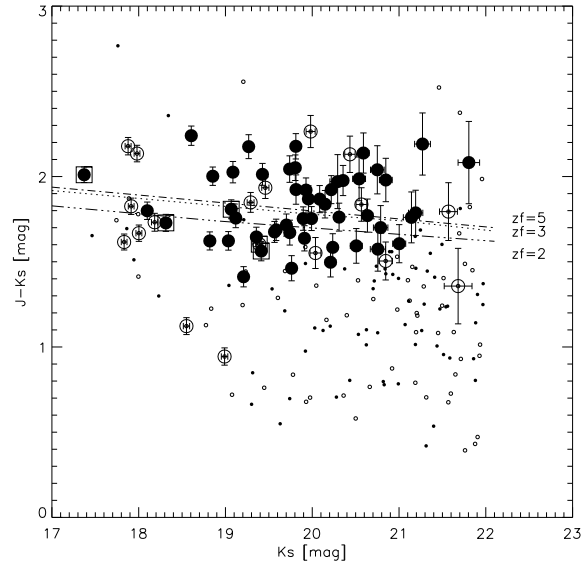}}}
\caption{$J$-$Ks$ versus $Ks$ colour magnitude relation. Squares mark spectroscopic members. Large symbols are photometric members, while small symbols are non members. 
Galaxies within 0.5 Mpc of the lensing galaxy are represented by filled symbols, while galaxies further away than 0.5 Mpc are represented by open symbols. 
The three lines are the predicted colour-magnitude relations at the cluster redshift ($z=1$) for passively evolving galaxies formed at $z_f$=2, 3 and 5 \citep{kodama97}.}
\label{JK}
\end{figure}

\begin{figure}
\setcounter{figure}{8}
\resizebox{\hsize}{!}{{\includegraphics{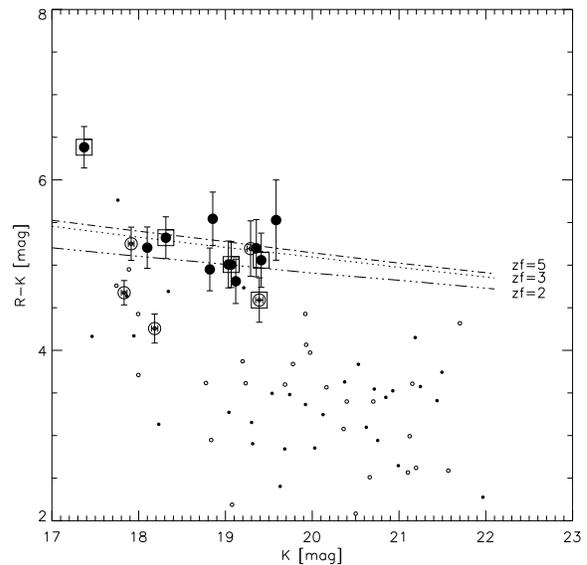}}}
\caption{$R$-$Ks$ versus $Ks$ colour-magnitude diagram for galaxies detected in the $R$ band. Symbols are as in Fig.~\ref{JK}}
\label{RK}
\end{figure} 

Most of the photometric members are concentrated in the ``inner region'' around the  lensing galaxy, while the non members are distribubuted more homogeneously across the field.
The photometric members  span a ``red sequence'' in the colour-magnitude diagrams with a considerable scatter.
The slope and zeropoint of the red sequence are hard to define due to the large scatter but are roughly consistent with predictions of the simple models.

The large observed scatter is likely to be caused by variations in the dust content, the morphology and/or the age of the stellar populations of the galaxies.
The red sequence is expected to increase its scatter and eventually fall apart as the redshift approaches the formation redshift of the stars in the galaxies.
The properties of the red sequence are consistent with the picture of a young cluster in the process of formation, in which the cluster galaxies habour young stellar populations and have not yet evoloved in to the old red galaxies found in the cores of relaxed clusters.

\subsection{Cluster galaxy luminosity function}
\label{LF}
We can now proceed to derive the $Ks$-band
luminosity function of the cluster galaxies. Since the $Ks$-band
observations are complete to $Ks=22$ and we know the photometric
redshift distribution of all galaxies in the field, we can do this
without having to make uncertain statistical corrections to account
for foreground and background contamination.
To take full advantage of the data, we use a maximum likelihood
technique applied directly to the luminosity distribution of the
cluster galaxies, rather than doing a ``least squares'' fit to a
binned representation. Details of the method are discussed in
App.\ref{lumfun}.  
In Fig.~\ref{contour} we show the best fitting Schechter function parameters
$\alpha=-0.60^{+0.39}_{-0.33}$ and $K^*=18.90_{-0.57}^{+0.45}$.
\begin{figure}
\setcounter{figure}{9}
\resizebox{\hsize}{!}{{\includegraphics{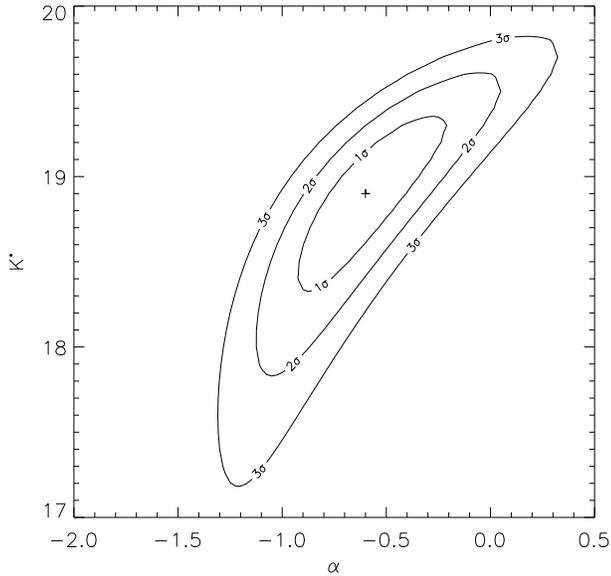}}}
\caption{Contour plot showing the  constraints on the  Schechter
function parameters derived from the maximum likelihood analysis. The
cross marks the best fitting parameters, the curves represents the
1-3$\sigma$ confidence levels  }
\label{contour}
\end{figure} 

In Fig.~\ref{lumfunfig} we plot the derived
$Ks$-band luminosity function of the cluster galaxies. 
It is found to be well represented by
the Schechter function.
\begin{figure}
\setcounter{figure}{10}
\resizebox{\hsize}{!}{{\includegraphics{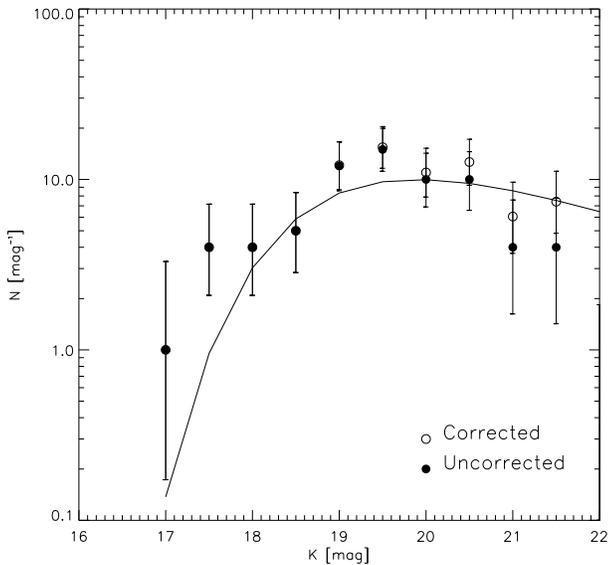}}}
\caption{$Ks$-band luminosity function of the cluster galaxies represented
by the Schechter function with the 
$\alpha=-0.60^{+0.39}_{-0.33}$ and $K^*=18.90_{-0.57}^{+0.45}$. The
filled symbols are a binned representation of the raw counts, while
the open symbols have been corrected for incompleteness using the
incompleteness function in Fig.~\ref{complfunct}. }
\label{lumfunfig}
\end{figure} 
The constraints on $\alpha$ are not strong, but it is important to leave it as a free parameter since the derived uncertainty on its value is coupled
to the derived value of $Ks^*$ and its uncertainties.
Also, since $\alpha$ has recently been shown to depend on wavelength \citep{goto02} it is interesting to note that the value of $\alpha$ is in agreement with
the value derived locally in the $z^*$-band (which roughly
corresponds to the same restframe wavelength as the $Ks$-band at
$z=1$) namely $\alpha_{z^*, local}=-0.58$.

Earlier studies of the  evolution of the luminosity function with
redshift have assumed the faint-end slope to be fixed at the value
measured in the Coma cluster $\alpha=-0.9$ \citep{depropris99,nakata01}.
To allow a comparison with these results we repeat our analysis with the faint-end slope fixed at
$\alpha=-0.9$ and derive a 1$\sigma$ constraint on the characteristic
magnitude of $K^*_s=18.60^{+0.22}_{-0.24}$.
In Fig.~\ref{kstarevol} we compare the derived value of $K^*_s$ with
values derived at other redshifts,
 and with the predictions of passively
evolving stellar population models with different formation redshifts
\citep{kodama97}. 
Dashed lines are calculated in a $\Omega_m=0.3$, $\Omega_{\Lambda}=0$, $h_0=0.6$   cosmology while full lines are calculated in a $\Omega_m=0.3$, $\Omega_{\Lambda}=0.7$, $h_0=0.7$ cosmology.
The filled circle is the value derived with $\alpha$ left as a free
parameter. The open circle 
is the value derived with a fixed $\alpha=-0.9$.
We have included this result in the plot to make a more direct
comparison with results from the literature. 
\begin{figure}
\setcounter{figure}{11}
\resizebox{\hsize}{!}{{\includegraphics{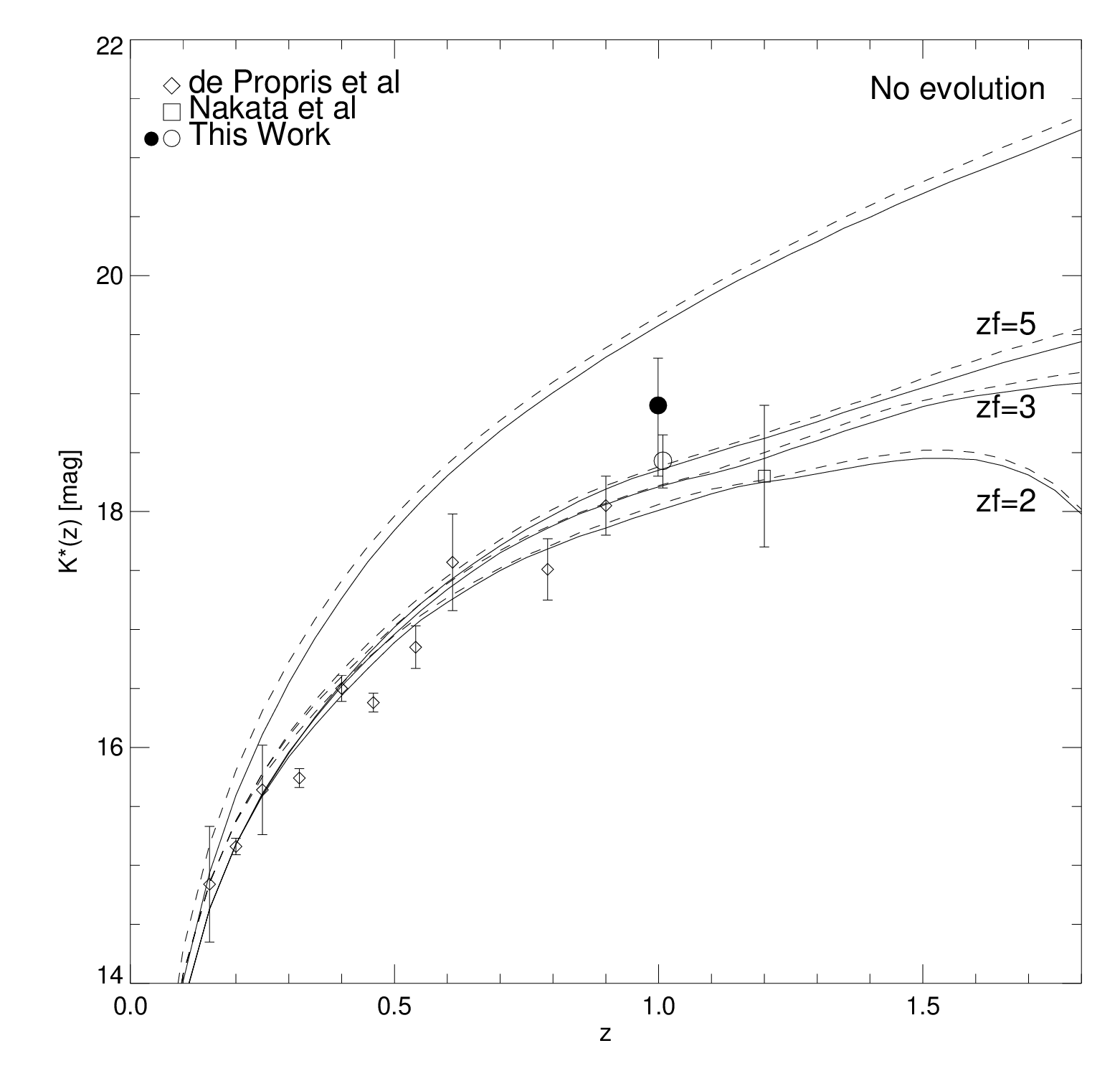}}}
\caption{Evolution of $K^*_s$ with redshift. 
The curves represent the
evolution predicted by stellar population synthesis models 
\citep{kodama97}, normalized to the Coma cluster which has $K^*_s=10.9$
\citep{depropris98coma}. Dashed lines are calculated in a
$\Omega_m=0.3$, $\Omega_{\Lambda}=0$, $h_0=0.6$   cosmology while the full
lines calculated in a $\Omega_m=0.3$, $\Omega_{\Lambda}=0.7$, $h_0=0.7$
cosmology. Lines are labeled according to the formation redshift of
their stellar populations. The lines labeled ``No-evolution'' only include
K-corrections.
The circles are the results of the present
work. The filled circle is the result derived with $\alpha$ as a
free parameter, while the open circle is the result obtained with
a fixed $\alpha =-0.9$.
The diamonds and the square are results from the literature  \citep{depropris99,nakata01}
}
\label{kstarevol}
\end{figure} 
As expected from Fig.~\ref{contour}, fixing  $\alpha$ results in a brighter best fitting $K^*_s$ with smaller errorbars,
bringing it into perfect agreement with the results of
\cite{depropris99} and \cite{nakata01} and consistency
with what is expected for a passively evolving stellar populations
formed at high redshift $z_f\ge 2$.  

To investigate the effects of field galaxy polution we redid the analysis on a subsample of the photometric member galaxies with photmetric redshifts in the range $0.9<z_{phot}<1.1$. This limits the effects of field galaxy polution but also excludes several  known (bright) cluster galaxies from the sample. The best fitting values obtained in this way are $\alpha=-0.51^{+0.43}_{-0.28}$ and $K^*_s=19.10_{-0.25}^{+0.77}$. 
The favored $K^*_s$ is a bit fainter, due to the exclusion of some of the brightetest galaxies from the sample, but apart from that the main effect is an increase of the size of the error bars.    
With the faint-end slope fixed at $\alpha=-0.9$, the best fitting characteristic magnitude hardly changes. The derived value is $K^*_s=18.55_{-0.27}^{+0.24}$.

From the present analysis we conclude that the effects of field galaxy polution on the derived luminosity function paramenters are marginal, but that fixing the faint-end slope of the cluster galaxy luminosity function directly affects the derived characteristic magnitude.
We find no evidence for a steepening or shift to fainter magnitudes of the
bright end of the (cluster galaxy) luminosity function at $z=1$, when compared to the local ($z=0$) one. Nor do we find a significant change of its faint-end slope to support the hierachical formation scenario.
\subsection{Mass-to-light ratio}
The restframe $K_s$-band mass-to-light ratio of the cluster around MG2016+112 can be estimated by comparing the total cluster mass, estimated from the velocity dispersion of 6  cluster members \citep{soucail01}, with the total K-corrected $Ks$-band flux of the cluster galaxies. 
We calculate the absolute K-corrected $Ks$-band magnitude of the cluster galaxies to be     
\begin{equation}
M_{Ks}=m_{Ks}-5\,{\rm{log}}\,d_L-25-k_{Ks}(z=1), 
\end{equation}
where $d_L$ is the luminosity distance in Mpc calculated in a $\Omega_m=1$,
$\Omega_{\Lambda}=0$, $h_0=0.5$ (for easier compariason to previous
results), and $k_{Ks}(z=1)=-0.352$ is the k correction in the $Ks$ band for an
object at redshift $z=1$ from \cite{poggianti97} (we apply this term
to be able to compare with the absolute $Ks$-band magnitude of the sun.
We assume $M_{V\sun}=4.8$ and $(V-Ks)_{\sun}=1.45$).

The derived $Ks$-band mass-to-light ratio of the cluster is found to be $M/L_{Ks}=27^{+64}_{-17}\, h_{50}\,(M/L_{Ks})_{\sun}$.
This is in agreement with the upper limit on the total mass of the cluster derived from X-rays \citep{chartas01} which when combined with our estimate of the total $Ks$-band luminosity of the cluster galaxies translates
into the following upper limit on $Ks$-band mass-to-light ratio
$M/L_{Ks}< 43\, h_{50}\,(M/L_{Ks})_{\sun}$, and in agreement
with the $Ks$-band mass-to-light ratio derived locally in the Coma
cluster $M/L_{Ks}=38\pm23\, h_{50}\, (M/L_{Ks})_{\sun}$ \citep{rines01}.

\section{Summary}
\label{summary}

We have obtained deep NIR and optical imaging of the field around the 
wide separation gravitational lens MG2016+112.
A photometric redshift analysis of objects in the field, reveal 69
galaxies with photometric redshifts consistent with being in a cluster
at the redshift of the lensing galaxy $z=1$.  
 
The $Ks$-band luminosity function of the cluster galaxies is well
represented by the Schechter function. The best fitting faint-end
slope $\alpha=-0.60^{+0.39}_{-0.33}$ is consistent with what is
measured at the same restframe wavelength (the $z^*$ band) locally, providing no evidence of evolution.
In the bright end no steepening or shift to fainter magnitudes of the
luminosity function is found. The best fitting characteristic
magnitude $K^*=18.90_{-0.57}^{+0.45}$ is consistent with what is
expected for a passively evolving population of galaxies formed at
high redshift $z_f\ge 2$.
From the $Ks$-band flux of the cluster galaxies and the total cluster mass estimates of \cite{soucail01} a  $Ks$-band mass-to-light ratio of  
$M/L_{Ks}=27^{+64}_{-17}\, h_{50}\, (M/L_{Ks})_{\sun}$ is derived, consistent with the upper limit derived from Chandra X-ray observations
\citep{chartas01} and the value measured locally in the Coma cluster \citep{rines01}.

The projected spatial distribution of the cluster galaxies is filamentary-like rather than centrally concentrated around the lensing galaxy, and show no apparent luminosity segregation. 
A handful of the cluster galaxies shows evidence of merging and interaction.
The cluster galaxies span a red sequence with a considerable scatter
in the colour-magnitude diagrams, suggesting that they contain young stellar populations in addition to the old populations of main sequence stars that dominate the $Ks$-band luminosity function. This is in agreement with spectroscopic observations which show that 5 out of 6 galaxies at the redshift of the lensing galaxy has emission lines.

The evidence presented here suggest that a young cluster of galaxies is in the process of assembling around the massive lensing galaxy MG2016+112. 
The cluster is mass-selected (from the precence of the wide-seperation gravitationally lensed QSO) and is therefore likely to be more typical of the $z\simeq 1$  cluster population than clusters selected using X-ray or optical techniques that rely on the presence of a relaxed intra cluster medium and/or a central concentration of old bright red galaxies for the clusters to be detected.

\appendix
\section{Deriving the Cluster Galaxy Luminosity Function using the
Maximum Likelihood technique}
\label{lumfun}
To take full advantage of the data, we  apply the maximum likelihood
technique of \cite{schechter} directly to the luminosity
distribution of the cluster galaxies, rather than doing a ``least squares'' fit to a binned representation.
In this way we include the information that in many  magnitudes intervals no galaxies are found, and we do not make the
 erroneous assumption of the $\chi^2$ method that the underlying
distribution is gaussian (it is in fact poissonian, either a galaxy is
there or it is not). This in turn leads to results with more realistic
errorbars.

In the magnitude interval
between $m$ and $m+dm$, the expected number of galaxies $p(m)dm$ 
is taken to be   
 
\begin{equation}
p(m)dm = f(m)n^* \left[ 10^{ 0.4(m^*-m) } \right ]^{\alpha+1}e^{-10^{0.4(m^*-m)}}dm,
\end{equation}
where $m^*$ is the characteristic magnitude of the cluster galaxies,
$\alpha$ is the slope of the faint-end of the cluster galaxy
luminosity function, $n^*$ is a normalization constant and $f(m)$ is the
completeness function of the ``photometric member'' sample (shown in
Fig.~\ref{complfunct}) evaluated in $m$.

Poisson statistics gives the probability that a cluster of
galaxies characterized by $n^*$, $m^*$ and $\alpha$ will yield the
data set actually observed ($N$; $m_1$,$m_2$,....,$m_N$; $m_{lim}$)
\begin{equation}
P={\rm{exp}}\left[ \int_{-\infty}^{m_{lim}} p(m)dm \right] \prod_{i=1}^{N}\left
[ p(m_i)dm \right], 
\end{equation}
where $m_{lim}$ is the magnitude limit (to which the dataset is
complete, in this case $m_{lim}=K_{lim}=22$).
The parameters $\alpha$ and $m^*$  which are ``most likely'' to
represent the observed data set are those that minimizes the C
statistics (the  likelihood function \citep{cash})
\begin{equation}
C=-2{\rm{ln}}(P).
\end{equation}
The minimum value of the C statistics $C_{min}$ is found by varying
$\alpha$ and $m^*$. 
For each combination of  $\alpha_i$ and $m_{*,i}$ we calculate the term $\Delta
C=C(\alpha_i,m_{*,i})-C_{min}$, which is distributed as $\chi^2$ with 2 degrees
of freedom \citep{cash}. From $\Delta C$ the  1-3$\sigma$
confidence levels can be derived in a straight forward manner as these
(as for the $\chi^2$ distribution) are found where $\Delta C = 2.29575, 6.18008, 11.8291$.

\label{lastpage}

\section*{Acknowledgments}
We thank T. Kodama for providing us with his elliptical galaxy evolution models, N. Drory and G. Feulner for letting us use their scripts for distortion corrections and H. Ebeling for useful discussions. This work was supported by the Danish Ground-Based Astronomical Instrument Centre (IJAF) and the Danish National Reseach Council (SNF).

\bibliography{2016paper}

\end{document}